# RADAR: A NOVEL FAST-SCREENING METHOD FOR

# READING DIFFICULTIES WITH SPECIAL FOCUS ON DYSLEXIA

Ioannis Smyrnakis[1], Vassilios Andreadakis[2], Vassilios Selimis[3], Michail Kalaitzakis[4], Theodora Bachourou[3], Georgios Kaloutsakis[4], George D. Kymionis[5], Stelios Smirnakis[6], Ioannis M. Aslanides[3]

[1]Technological Educational Institute of Crete, Greece;

[2]Optotech Ltd., Crete, Greece;

[3]Emmetropia Eye Institute, Crete, Greece;

[4]Medotics Ltd., Basel, Switzerland;

[5]Athens Medical School, University of Athens, Greece and Medical School, University of Lausanne, Jules Gonin Eye Hospital, Switzerland;

[6]Department of Neurology, Brigham and Women's Hospital and Jamaica Plain VA Hospital, Harvard Medical School, Boston, Massachusetts, USA.

Corresponding Author
Ioannis M. Aslanides
Heraklion, Crete, Greece
Email: i.aslanides@emmetropia.gr

# Abstract

Dyslexia is a developmental learning disorder of single word reading accuracy and/or fluency, with compelling research directed towards understanding the contributions of the visual system. While dyslexia is not an oculomotor disease, readers with dyslexia have shown different eye movements than typically developing students during text reading. Readers with dyslexia exhibit longer and more frequent fixations, shorter saccade lengths, more backward refixations than typical readers.[1,2] Furthermore, readers with dyslexia are known to have difficulty in reading long words, lower skipping rate of short words, and high gaze duration on many words. It is an open question whether it is possible to harness these distinctive oculomotor scanning patterns observed during reading in order to develop a screening tool that can reliably identify struggling readers, who may be candidates for dyslexia.

Here, we introduce a novel, fast, objective, non-invasive method, named Rapid Assessment of Difficulties and Abnormalities in Reading (RADAR) that screens for features associated with the aberrant visual scanning of reading text seen in dyslexia. Eye tracking parameter measurements that are stable under retest and have high discriminative power, as indicated by their ROC (receiver operating characteristic) curves, were obtained during silent text reading. These parameters were combined to derive a total reading score (TRS) that can reliably separate readers with dyslexia from typical readers. We tested TRS in a group of school-age children ranging from 8.5 to 12.5 years of age. TRS achieved 94.2% correct classification of children tested. Specifically, 35 out of 37 control (specificity 94.6%) and 30 out of 32 readers with dyslexia (sensitivity 93.8%) were classified correctly using RADAR, under a circular validation condition (see section Results/Total Reading Score) where the individual evaluated was not included in the test construction group.

In conclusion, RADAR is a novel, automated, fast and reliable way to identify children at high risk of dyslexia that is amenable to large-scale screening. Moreover, analysis of eye movement parameters obtained with RADAR during reading will likely be useful for implementing individualized treatment strategies and for monitoring objectively the success of chosen interventions. We envision that it will be possible to use RADAR as a sensitive, objective, and quantitative first pass screen to identify subjects with reading disorders that manifest with abnormal oculomotor reading strategies, like dyslexia.

# INTRODUCTION

Reading is a complex skill that improves with experience, and becomes increasingly automatic and accurate in early childhood. Nevertheless, some subjects, despite having above average IQ as well as healthy vision and hearing, struggle to become fluent readers, and have great difficulty in fully understanding written text. These subjects, if not diagnosed with another well described syndrome (e.g. ADHD), are frequently labelled as dyslexics. According to Lyon et al. "Dyslexia is a specific learning disability that is neurobiological in origin. It is characterized by difficulties with accurate and/or fluent word recognition and by poor spelling and decoding abilities. These difficulties typically result from a deficit in the phonological component of language that is often unexpected in relation to other cognitive abilities and the provision of effective classroom instruction".[3] As individuals grow older, compensating mechanisms develop that help alleviate the symptoms of dyslexia.[4] However, the learning gap that has developed by then follows dyslexic readers for much longer.[5]

During childhood, reading disabilities of the dyslexia type, are the most common learning disabilities, and influence student performance in and out of school. Reading disabilities can negatively impact student academic achievement, self-image, social adaptation, as well as societal attitudes towards affected children. Timely identification and diagnosis of reading disabilities is important to guide intervention and to avoid personal, academic, and social repercussions. Hence, early detection of dyslexia, and other similar reading difficulties, is imperative in order to provide the necessary help to students.

Identification of students with dyslexia is not usually made until grade 3 of elementary school, when reading ability lags behind what is expected for age, and starts to hinder overall educational progress. In fact, identification is often made much later, or even worse, "significant numbers of students with dyslexia go undiagnosed and their symptoms unaddressed, with tragic results".[6] Full diagnosis of dyslexia is an elaborate process, which, according to the International Dyslexia Association[7], has to consider multiple factors including: background information, intelligence, oral language skills, word recognition, decoding, spelling, phonological processing, reading comprehension, to name a few. Although full diagnosis of dyslexia is clearly essential for therapeutic treatment, it is often not practical for screening large populations of school-age children since it entails long delays[8] and significant cost.[9,10] In Greece, waiting lists for evaluating children suspected for dyslexia can be as long as 14 months (waiting list, Center for Differential Diagnosis, Diagnosis and Support, department of Heraklion, December 2014), and this experience is not unique to this country. In Ireland, an evaluation could take up to 9 months.[8] Although relatively fast screening tests do exist (e.g. Dibels in US), they almost exclusively depend on loud reading, which involves the double task of decoding and phrasing text, and furthermore they do not make use of the visual attention data, that eye tracking may provide. Developing a rapid screening method, based on eye-tracking technology, with high sensitivity and high negative predicting value for detecting reading disabilities would be of considerable benefit.

Eye movements are characterized by fixations and saccades. Fixations are the intervals during which the eye remains still (barring micro-saccades) and they typically last 200-300 ms. Saccades are very fast (up to 500°/sec) movements of the eye that occur between consecutive fixations. Several studies have analyzed eye movements of participants during reading.[1,11] The main objective of these studies was to elucidate the underlying cognitive processing that occurs during reading. It has been hypothesized that saccades guide the eyes to obtain the maximum possible information from the visual scene, and this strategy may be used in reading.[12] The average fixation duration during reading in typical adult English readers is 200-250ms and saccades have an average length of 7-9 characters.[13] Note that the appropriate unit to use for saccade length is the number of characters, since the number of letters traversed during reading is relatively independent of the viewer's distance (for the same text), even though the letter spaces subtend different visual angles.[14] For children 7-11 years old, fixations range mostly from 243-285ms and the average saccade length is 7-8 characters.[11] Overall, the pattern of eye movements during reading depends on multiple factors, including basic oculomotor control mechanisms, attention, meaning (semantics), and the syntactic structure of a text.[15,16]

It should be mentioned at this point that although dyslexia is not a primary oculomotor problem, eye movements differ during reading between typical and dyslexic readers.[1,2] Multiple studies[17,18,19,20] have shown differences in the eye movements of typical readers versus readers with dyslexia. Typically, in readers with dyslexia, fixation duration and number of fixations increase, average saccade length gets shorter and the number of regressions (short backward eye movements targeting text that has already been read) increases.[1,2] The observed differences can be attributed to abnormal linguistic or cognitive processing. Several possible etiologies have been proposed, such as left-right visual field imbalance, sluggish attention or "crowding".[21] For example, when words are flashed on either side of fixation, typical readers show a bias for word recognition favoring the right visual field while dyslexics favor the left.[22] This is thought to cause "crowding" on the left of fixation during reading in dyslexic subjects. In addition, some dyslexics exhibit a delay in shifting attention resulting in longer fixations,

"crowding," and backward refixations during reading[21]. It is important to note that, even though dyslexia is not a primary oculomotor problem[1,2], it may be possible to identify readers with high probability of dyslexia by harnessing the distinctive oculomotor scanning patterns observed in this population during reading.[21]

The RADAR method we propose is a method that is based on eye tracking technology to evaluate silent reading. It uses two types of parameters to evaluate a subject. The first type, termed non word specific, consists of the fixation durations, the saccade lengths, the short refixations (less than 4 characters) and the total number of fixations during reading. The second type, termed word specific, is based on the gaze duration (overall fixation duration at first visit) on each word and the number of revisits on each word. It then produces a score, which allows the subject to be classified with high probability as dyslexic or not. RADAR, as a dyslexia screening method, has the advantage of producing an objective, rapid and reproducible initial assessment that can be used to quickly screen school age children, leading to the early identification of participants that need more extensive testing and intervention.

Advantages of RADAR as a screening method over previous methods, include: 1) Deriving and using a rich set of oculomotor data collected during reading, ii) evaluating subjects during silent reading, which is typically not assessed during standard printed tests (e.g. DIBELS), and 3) is able to ascertain errors of reading associated by lack of effort by analyzing the scanpath displayed by the eye-tracker during reading. We argue below that one reason for the value of RADAR as a screening test is that children with a negative RADAR result do not have dyslexia with high probability. If used in a population where dyslexia prevalence is 10% a child with a negative RADAR result has 99.3% chance of being non-dyslexic (see Discussion). In this way, a small fraction of children with positive RADAR result can be identified early and referred for a full diagnostic procedure. The referred fraction of children has high probability to have a reading disorder, which is most likely dyslexia but may also include other syndromes that affect reading. The exact identity of the reading difficulty could be suggested by the nature of the RADAR results but would be clarified during the full diagnostic procedure. This would reduce the volume of work in the dyslexia diagnostic centers and hence reduce both the cost of diagnosis and the size of the waiting lists.

It should be noted that diagnostic characteristics of dyslexia may differ in different languages. For example, several studies[18,23,24,25] concluded that in transparent languages (regular orthographies), such as Italian, German, and Greek, the phonological deficits of the individuals with dyslexia tend to be less pronounced compared to non-transparent languages, such as English. The main reason for this difference seems to be that for the transparent languages prior knowledge on how to pronounce a series of letters is required only infrequently, and by default the grapheme–phoneme rules are well correlated to phonological forms.[26] Despite the research studies that have been performed regarding the reading characteristics of typical and/or dyslexic readers using eye-tracking, it remains an open question whether eye movement patterns during reading can be used to generate a reliable screening tool that can separate readers with dyslexia from typical readers in different languages.

In this manuscript, we show that it is in fact possible to use eye movement patterns and characteristics in order to differentiate 8.5-12.5 years old children with dyslexia from those without dyslexia in the Greek language (a preliminary study in the English language is also presented). We use eye movement parameters as a benchmark to classify children as either potentially dyslexic readers or as typical readers. Although the RADAR method cannot yet be used as a definitive diagnosis tool by itself that can distinguish among different syndromes that cause reading difficulties, it can quickly screen large populations of children to pick a small subgroup that has high probability of having dyslexia. Children not picked by RADAR fit well with the typical reader characteristics and as such are not likely to have reading difficulties of any kind. In this study, we evaluate the reading process in young Greek speaking children (69 children aged 8.5-12.5). We focus on deriving measures that allow us to discriminate between typical (normal) readers and readers that exhibit difficulties that are consistent with dyslexia. We combine these parameters into a score that is used to discriminate readers with dyslexia and demonstrate that this score is effective in discriminating readers likely to have dyslexia from typical readers. Importantly, we replicated the study with a group of English speaking participants, confirming that our method can be applied not only to transparent languages, like Greek, but also to non-transparent languages, like English.

Having a fast, automated and reliable way to identify children at risk of dyslexia and potentially other reading abnormalities is a significant advance, particularly if proved amenable to large-scale screening. If used properly, RADAR has the potential to help identify young children with possible dyslexia after grade 2 of primary school in a fast and simple way, leading to potentially faster intervention.

## METHODS

### Participants

Seventy-eight children (8.5-12.5 years old; 42 girls and 36 boys) participated in the study with native Greek speaking children in Greece. Nine of the participating children (3 control, 6 dyslexic) were rejected blind to the diagnosis due to unreliable eye-movement recording or lack of cooperation with the experimenters. None of the participants was rejected due to poor vision or hearing. Of the remaining 69 children, 32 (15 girls, 17 boys) were diagnosed as dyslexic by the official governmental agency for diagnosing learning and reading difficulties in Greece. The evaluation procedure for the dyslexia diagnosis included social and educational background evaluation, and Test-A for word decoding, fluency, syntax and text comprehension.[27] The diagnosis was delivered by a committee consisting of a social worker, a psychologist and a special education teacher, evaluating all these factors. In addition, the IQ of the participants was evaluated by a psychologist (WISC III normalized for Greek population). Contrary to many western countries, no reading tests are administered to Greek children within the school system, hence no reading test scores were available for the participants. These children constitute the dyslexic group of our study. The remaining 37 children (22 girls, 15 boys) did not have reading difficulties. They were recruited at random among children that didn't exhibit reading problems at school. These children were furthermore assessed by a special education needs teacher to be without any reading or learning difficulty. These children constitute the control group of our study. All children were between third and sixth grade in the Greek primary school system, and they turned out to be 8.5-12.5 years of age. All children were native Greek speakers and underwent a session with the RADAR method.

To participate in the study, every child had to have IQ score above 90[28,29] (in Greece, 80-89 is considered low normal and children with this score were not included in the study because it was necessary to have reasonably developed reading skills even for children 8.5 years of age). For the Greek population, the mean IQ score is 100 and the standard deviation is 15.[30] Also, normal visual acuity (with or without correction) and normal hearing was required. No other inclusion criteria were applied. All children were assessed for ophthalmological problems by an ophthalmologist participating in the study, blind to the dyslexia diagnosis. Furthermore, no children with pronounced hearing problems participated in the study. After a month, a retest phase was initiated with a subgroup that had been tested in the first trial. The rationale behind this was to evaluate the test-retest correlation of the RADAR parameters. Thirty-five children (17 girls, 18 boys) were re-tested in the same text. Of those, 13 children (7 girls, 6 boys) were in the control group (typical readers) and 22 children (10 girls, 12 boys) were in the dyslexic group.

There is in general a positive correlation between IQ and reading ability for up to 12[th] grade school children in both control and dyslexic populations. However, reading ability and IQ develop in parallel with age in control population while this is no longer true in dyslexic readers. In dyslexic readers IQ develops faster than reading ability with age. This indicates a difference in the development of cognition and reading.[31]

Note that the experimenters were blinded regarding the diagnosis of the children that were assessed, prior to producing and analyzing the RADAR results.

**Materials**

Participants were asked to silently read a text in the Greek language, presented on a computer monitor, while their eye movements were recorded. The instructions given to participants were to read the whole text, at their own pace, that they did not need to rush their reading since the purpose of the task was not to read fast, and that at the end of the reading they would have to answer to five comprehension questions. The text (called basic text below) was written by a special education teacher in order to be appropriate for the participants' age group (8.5-12.5 years). The text had 181 words, many of which were multi-syllable. The statistics of the text are shown in Table 1.

**Table 1. Text statistics for the basic text.**

| | |
|---|---|
| Total word count: | 181 |
| Unique words: | 114 |
| Total number of characters: | 1168 |
| Number of characters without spaces: | 986 |
| Average characters per word: | 5.44 |
| Average syllables per word: | 2.37 |
| Sentence count: | 13 |
| Max sentence length (words): | 8 |
| Min sentence length (words): | 2 |

Various text metrics for the basic text.

Text stimuli were presented in font Courier New, mono-spaced, font size 30pt, bold, line space 2.3 lines, black colored on grey background, on an 18.5-inch flat-panel monitor at 50cm distance from the participants' eyes. Display resolution of the monitor was set to 1366 × 768 pixels with a refresh rate of 60 Hz. At the distance of the chin rest from the screen (approx. 50 cm), the horizontal width of the screen corresponds to 46.4 degrees and the vertical width to 26.1 degrees, nevertheless the text was confined to approx. 30 degrees to avoid eye-tracking loss. The text had 28 lines and it was divided in five screens, hence 6 lines were presented on each of the four screens and 4 in the last one. At the beginning of the experiment, the individual was instructed to press the "Space Bar" key when finished reading the first screen in order to move to the next one etc.

Five comprehension questions were asked after the full reading of the text and were answered orally with a "YES" or "NO" and they were automatically stored in the database. Answers were scored with: 0 points for missing or incorrect answers and 1 point for a correct answer, resulting in a possible maximum comprehension score of 5 points. The purpose of these questions was to maximize the chance that the participants read the text and not to enforce understanding. That is, the number of correct answers was not used to reject participants and was not part of the RADAR evaluation. The reading path of the eyes (see Fig. 1 for stimuli and Fig. 2 for reading path) ensured that the text was in fact read.

**Apparatus**

For conducting the experiments, we developed a custom-made eye-tracker. Both software and hardware have been designed and manufactured by Medotics AG. The motivation behind building our own tracker instead of using a commercial device are the following: i) create a device that will be user friendly for the children, ii) extract accurate results, iii) perform experiments and data analysis in a short time, iv) result verification (commercial trackers do not give access to the way they compute specific eye tracking parameters, e.g. fixations). To be able to achieve the above we looked into the following features: i) matched the device's functional specifications with the physical dimensions the participants have, ii) created an easy-to-use graphical interface, ii) enabled a two-function operation (during the experiment a technician is supervising the process to ensure result accuracy), iv) combined system calibration, eye tracking experiments and result generation under the same software platform.

The tracker consists of two steady cameras that can record images up to 60Hz with a resolution of 1600x1200 pixels. While the participant performs a reading task, the cameras record the participant's face. Cameras are positioned between the screen and participant with a viewing field from down towards the participant's face. The specific setup improves visibility of participants wearing glasses vs if the cameras have been placed on the top of the screen. The images extracted are then used to detect pupil and corneal reflection (CR) coordinates. The corneal reflections are caused by an infrared light source. The source is a bundle of different LED emitters that were installed in a way to minimize further reflections or shadows to the participant's face. Eigenfaces software is used for computing the pupil/CR positions. The approach does not require a colored image and thus the cameras installed have a monochrome CMOS image sensor.

The experiments are conducted in the following way (see Fig. 1). The individual is placed in front of a computer screen and (s)he is asked to place his/hers head on a chinrest. The chinrest minimizes head movements and head tremor improving accuracy significantly. Once the participant is settled the calibration process begins. A series of 25 different spots (points) are displayed consecutively. The spots are being projected on known coordinates symmetrically positioned on a matrix grid. We have chosen their positions in a way that they evenly cover the screen to ensure accurate calibration over the whole range of eye movements. The participant is asked to fixate at the displayed spots. Each spot is being displayed until enough saccades are being recognized in its region.

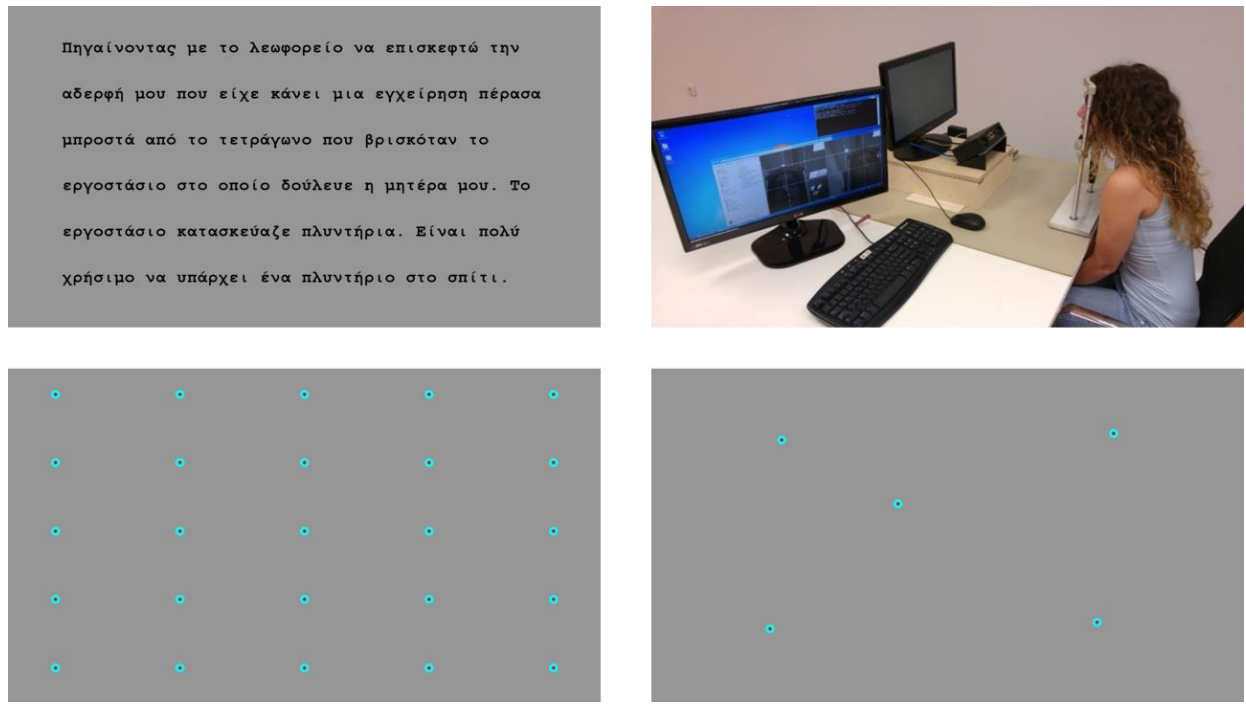

**Figure 1:** _Upper left_: A screen from the stimuli. The text was split in 5 screens and had 181 words. _Upper right:_ A participant is seated in front of the monitor. With the support of a chin rest she is reading the text displayed. _Bottom left:_ 25-point calibration screen. _Bottom right:_ 5-point validation screen.

When the calibration finishes, we map the pupil and CR coordinates on the camera images to the coordinates of the spots that the participant was looking. After the mapping is complete we proceed to the validation process. We display 5 new points on the screen (Fig. 1, bottom right) and we measure the accuracy of the mapping from the previous step. The point coordinates have been chosen now in a rather less geometrical way. They are positioned on places where later a text to be read will be displayed. An accuracy threshold is being chosen automatically, and once passed, the reading can start, otherwise the calibration process starts again. This procedure ensures that we have an accurate calibration and high quality data for analysis. Following initial calibration, the position of the words in the text serves to validate the calibration at the end of reading.

During the experiment both eyes are visible from each camera. The calibration process produces two different matrices for each eye. The first matrix is for the CR distance from the pupil center and the second matrix for their relative angle. In total, we have 8 matrices (2 cameras X 2 eyes X 2 matrices-per-eye). The matrices are being computed by linear interpolation of the measurements obtained during calibration. The dimension of the matrices is about half of the number of the screen display resolution in pixels. For example, if the screen has resolution of 1366 pixels by 768 pixels, then the matrices have 685 columns by 385 rows. For each eye that we have a pupil and CR detection we measure the length and the angle of their distance. At each frame then we compute the matrix value, which has the minimal divergence from the distance values (length and angle). The gaze is then computed linearly based on the minimal divergence described above.

A critical parameter of the eye tracking analysis is to identify and estimate the position and duration of fixations. Fixations display where the participant's reading is at, since they represent localized "clouds of gaze points". We use a dispersion algorithm[32] to identify fixations. It is based on two thresholds, minimum fixation duration ($t_{min}$) and maximum gaze point scatter, $d_{max}$. To constitute a fixation, gaze points have to remain within a window of size $d_{max}$ for at least $t_{min}$. Subsequent gaze points belong to the same fixation cloud as long as they remain inside the $d_{max}$ window. These gaze points define a fixation with position the center of the cloud and duration the time that gaze points remain inside the $d_{max}$ window. In our analysis, we empirically determined that $t_{min} = 90ms$ and $d_{max} = 45pixels$, corresponding to 1.5°, work well for identifying fixations across all participants (Fig. 2). Blinks are identified (in general 4-5 blinks per minute) and excluded from the analysis.[33] Fig. 2 shows the recognized fixations in the reading path of a typical reader (left) and a reader with dyslexia (right).

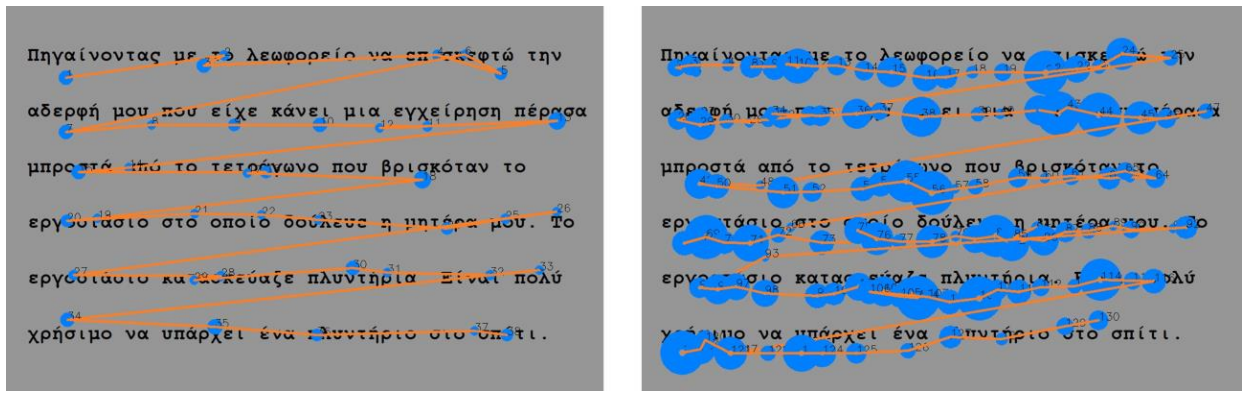

**Figure 2:** A reading path from a typical (normal) reader (_left_) and from a reader with dyslexia (_right_). The blue circles are the fixations and the orange lines are the saccadic movements. The bigger the circle, the longer the fixation. Clearly the reader with dyslexia exhibits longer fixations duration, shorter saccadic movements, regressive movements and longer reading time than the typical reader.

## Procedure

In both groups, children were tested individually in a quiet room, after a parent/guardian signed a written consent form. Two members of the RADAR's team conducted the experiment and they were inside the room for the duration of the session. Before the reading assessment, a basic vision screening was performed by an ophthalmologist. The IQ testing was performed by a psychologist. The IQ test used was the Greek normalized version of the WISC test. A participant was excluded from the study if the IQ score was below 90 (The reason for this choice is twofold. First it was desired to have reasonably developed reading skills even for children 8.5 years of age, and second, the usual approach to dyslexia in Greece is that a child with IQ less than 90 cannot be labelled as dyslexic.[34] In Greece, IQ 80-89 is considered low normal. Note that mean IQ in Greece is 100, and its standard deviation is 15).[30] If the participants passed the vision screening (here all participants passed the vision screening test), then they were seated in front of the presentation monitor and received the directions for the upcoming task on the display screen in front of them. To minimize head movements, participants were asked to place their heads on a chin rest. If the members of the RADAR team noticed that the child was struggling to hear the instructions (i.e. abnormal hearing), then the participant was rejected. Overall, 9 children were rejected, due to unreliable eye-movement recording or lack of cooperation with the experimenters. The experimenters explained the procedure aloud to the children, describing each step, to ensure that they understood what they had to do. The text comprehension questions were only presented orally to the participant after the text was read. Participants received identical directions for the reading task, instructing them to read the text so that they were able to answer comprehension questions. A calibration procedure preceded the reading task. The individual was instructed not to move from their position and not to talk with the experimenters while reading. A testing session typically lasted approx. 20mins, including the transfer back and forth from the classroom.

## ANALYSIS AND RESULTS

The two synchronized 60Hz cameras of the eye tracker recorded the point of gaze from the pupil center and the corneal reflection of both eyes. One point of gaze was recorded every 1/60 seconds. These gaze points were clustered to fixations with a standard moving window dispersion algorithm, with maximum dispersion threshold 1.5 degrees and time threshold of 90ms, as described in the Methods (Apparatus). The length of a saccadic movement is calculated along the x-axis, as the difference between the x coordinates of the centers of consecutive fixations. The speed criterion applied for saccade recognition was that saccade speed be >100°/sec. The relatively low speed threshold chosen was due to the 60Hz frequency of the cameras, which is too low to evaluate higher saccade speeds with accuracy. Since the minimum time interval that can be measured is 16ms, the maximum speed quotient we can assign to a 1.65° degree saccade (spanning 2 characters) is 1.65/0.016 = 103° /sec. Note that the actual saccadic speed can be much larger because the 1.65° movement can occur in less than 16ms[32], however the threshold chosen works well for our analysis. For our analysis, it was reasonable to put a lower limit on the saccade speed at 100°/sec in order to exclude intra-fixation dispersion (micro-saccades are difficult to detect reliably under our conditions). Two types of fixation analysis followed: i) General (non word-specific) analysis, and ii) word-specific analysis.

**General, non word-specific, analysis**

The parameters analyzed were: i) fixation duration, ii) saccade length, iii) the short backward saccades (less than 4-character-long refixations representing mostly within-word backwards eye movements), and iv) the total number of fixations during the reading of the text. We call these parameters general (non word-specific) because the fixations are not associated with specific words in the text. Since in this analysis only the relative position of the fixations matters, a global shift of the fixation pattern does not affect the general analysis.

To analyze the fixation duration, we fit an Exponential-Gauss (Exp-Gauss) distribution using the probability density function $f(x; \mu, \sigma, \lambda) = \frac{\lambda}{2} e^{\frac{\lambda}{2}(2\mu + \lambda\sigma^2 - 2x)} erfc\left(\frac{\mu + \lambda\sigma^2 - x}{\sqrt{2}\sigma}\right)$, to the fixation duration frequencies. Here $erfc$ is the complementary error function defined as $erfc(x) = 1 - \mathrm{erf}(x) = \frac{2}{\sqrt{\pi}} \int_x^{\infty} e^{-t^2} dt$. The Exp-Gauss distribution, which is the convolution of a Gaussian and an Exponential distribution, has been found to be appropriate for describing the fixation duration frequencies.[35,36] The parameters that describe the Exp-Gaussian distribution are μ and σ, the mean and standard deviation of the Gaussian part of the distribution, and $\tau = 1/\lambda$, the parameter describing the exponential decay of the Exponential part of the distribution. The analysis of the significant fixation duration parameters is shown in Table 2.

To analyze the saccade length, we evaluate the quartiles $q_{25\%}$, $q_{50\%}$, $q_{75\%}$ of the frequency distribution of the saccade lengths.[37] The next parameter evaluated was the number of backward refixations ($R$) that resulted from backward saccades of length below an appropriately chosen threshold, during reading of the particular text. To determine the optimal backward saccade length threshold for the desired classification, we evaluated each threshold through the ROC curve area of the number of backward refixations. A threshold range between 100 and 400 pixels (3.4°-13.5° or 4-16 text characters) was evaluated (Fig. 3). The analysis of the significant saccade length parameters and the refixations parameter are shown in Tables 3 and 4.

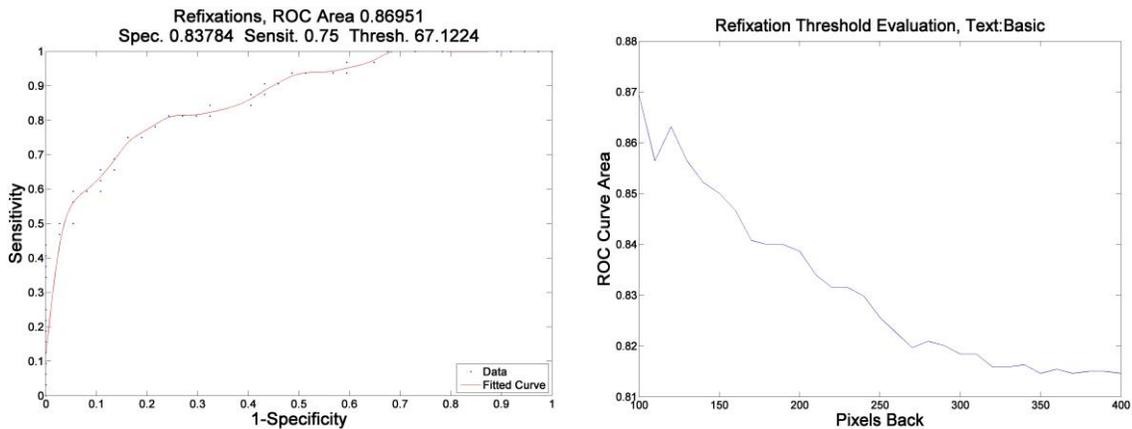

**Figure 3:** _Left_: The ROC curve for backward refixations of less than 100 pixels (4 characters). _Right:_ On the x-axis we have the threshold used to count refixations. This number of refixations is used to derive the ROC curve of the parameter, which is used as a classifier. On the y-axis we have the area under this ROC curve. As we can see, we get maximum area at a threshold of 100 pixels, which corresponds to 4 characters. The fitting is done with smoothing splines. Each data (blue) point corresponds to a sensitivity-specificity pair associated to a range of values for the threshold.

It is clear from these curves that the optimal backward saccade threshold is 100 pixels (4 characters) for this text. So, backward refixations correspond to backward saccades of less than 100 pixels (3.4°). The threshold was not taken to be below 100 pixels (4 characters) because there are too few refixations below this threshold, and because of precision considerations.

The final general parameter that was evaluated was the fixation number (Fix) during reading of the whole text. The analysis of the fixations number parameter is shown in Table 5.

**Word-specific analysis**

The second type of fixation analysis we performed, was the word-specific analysis. This analysis was performed on the first pass from each word. Hence, the gaze duration on each word starts upon first visit on the word and finishes when the eye leaves this word for the first time. The first set of parameters measured in this analysis was the number of skipped words $F^0$, the number of singly fixated words at first pass $F^1$, and the number of multiply fixated words at first pass $F^{\geq 2}$. Next, we evaluate the performance of the gaze duration on words of varying length measured in characters. For each word, the gaze duration of the control population is used to construct a Gaussian distribution curve. The gaze duration of an individual is placed on this Gaussian curve, and a score is given on the particular word by the cumulative probability corresponding to his gaze duration. The number $X_L$ of words of length L (in characters) with score above 90% is a surrogate marker used to mark the difficulty that the participant has in reading words of a particular length. The second set of parameters is based on $X_L$, where L represents word lengths appropriate to the particular text. For our basic text, parameters used are $X_{L \leq 3}$, $X_{L=4,5}$, $X_{L=6,7}$, and $X_{L \geq 8}$.

**Parameter Analysis**

For all parameters, the following evaluation procedure is followed: First the area under the ROC curve, $A_{ROC}$, is measured. The closer $A_{ROC}$ is to 1, the better the parameter in discriminating among the two populations, irrespective of the threshold one uses. The threshold can be chosen later according to the tradeoff between sensitivity and specificity that is required by the test (see later on Fig. 6). Parameters with $A_{ROC} \geq 0.87$ were retained. Second, we compare the difference of each parameter mean between the control and the dyslexic populations ($|\overline{P_{Contr}} - \overline{P_{Dysl}}|$) with the test-retest variability within each population ($\overline{|P_{Test}^{Norm} - P_{Retest}^{Norm}|}$ and $\overline{|P_{Test}^{Dysl} - P_{Retest}^{Dysl}|}$). For this text, to retain a parameter we require that $|\overline{P_{Contr}} - \overline{P_{Dysl}}|$ is at least twice as big as the larger of $\overline{|P_{Test}^{Norm} - P_{Retest}^{Norm}|}$ and $\overline{|P_{Test}^{Dysl} - P_{Retest}^{Dysl}|}$. Retained parameters had to satisfy both of the above criteria.

The parameters that were retained are the following: The τ parameter of the fixation duration distributions, the three quartiles $q_{25\%}$ $q_{50\%}$, $q_{75\%}$ of the saccade length distribution, the number of backward refixations below 4 characters $R$, the number of fixations $Fix$, the number of skipped words $F^0$, the number of multiply fixated words at first pass $F^{\geq 2}$ ($F^1$ is omitted since it is fully dependent on the other two), and $X_{L\leq 3}$, $X_{L=4,5}$, $X_{L=6,7}$, and $X_{L\geq 8}$ for our text.

**Fixation Duration:** The parameter τ was retained, since it gave $A_{ROC} = 0.91$ for the text and it satisfied the criterion of stability under test-retest. Parameters μ and σ were rejected. Here $r_\tau$ is the Pearson correlation of the test-retest values of $\tau$ for the combined Control-Dyslexic population, showing excellent test-retest consistency. See data on Table 2.

**Table 2. Parameter τ for the basic text.**

|  | $A_{ROC}$ | $\overline{\|\tau_{Test}^{Contr} - \tau_{Retest}^{Contr}\|}$ ±SD | $\overline{\|\tau_{Test}^{Dysl} - \tau_{Retest}^{Dysl}\|}$ ±SD | $\overline{\tau_{Contr}} - \overline{\tau_{Dysl}}$ | $r_\tau$ |
|---|---|---|---|---|---|
| τ $_{bacic}$ | 0.91 | 13.27ms ±12.22 | 27.78ms ±23.88 | -91.3ms | 0.93 |

The analysis of the parameter τ.

**Saccade Length:** The data obtained is summarized in Table 3.

**Table 3. Saccade length for the basic text.**

|  | $A_{ROC}$ | $\overline{\|q_{Test}^{Contr} - q_{Retest}^{Contr}\|}$ ±SD | $\overline{\|q_{Test}^{Dysl} - q_{Retest}^{Dysl}\|}$ ±SD | $\overline{q_{Contr}} - \overline{q_{Dysl}}$ | $r_q$ |
|---|---|---|---|---|---|
| $q_{25\%}$ | 0.96 | 15.87pxl ±16.11 | 6.29pxl ±4.69 | 34pxl | 0.90 |
| $q_{50\%}$ | 0.96 | 21.00pxl ±16.18 | 9.25pxl ±7.82 | 56.1pxl | 0.95 |
| $q_{75\%}$ | 0.92 | 29.65pxl ±22.63 | 14.59pxl ±17.83 | 75.8pxl | 0.91 |

The analysis of the saccade length.

Here all quartiles are retained, however they are not treated as independent parameters in deriving the TRS (see Appendix).

**Refixations:** The results for the number of refixations $R$ is given in Table 4.

**Table 4. Refixations for the basic text.**

|  | $A_{ROC}$ | $\overline{\|R_{Test}^{Contr} - R_{Retest}^{Contr}\|}$ ±SD | $\overline{\|R_{Test}^{Dysl} - R_{Retest}^{Dysl}\|}$ ±SD | $\overline{R_{Contr}} - \overline{R_{Dysl}}$ | $r_R$ |
|---|---|---|---|---|---|
| $R_{basic}$ | 0.87 | 9.92 ±8.27 | 17.55 ±14.31 | -45.5 | 0.88 |

The analysis of the refixations.

The number of refixations was retained. Although its $A_{ROC}$ is below 0.9, we use it as a parameter because it can be associated to lexical difficulty, and as such it is particularly important in the diagnosis of dyslexia.[13,38]

**Fixation Number**: The results for the number of fixations $Fix$ is summarized in Table 5.

**Table 5. Fixations Number for the basic text.**

| | $A_{ROC}$ | $\left|\overline{Fix_{Test}^{Contr} - Fix_{Retest}^{Contr}}\right|$ ±SD | $\left|\overline{Fix_{Test}^{Dysl} - Fix_{Retest}^{Dysl}}\right|$ ±SD | $\overline{Fix_{Contr}} - \overline{Fix_{Dysl}}$ | $r_F$ |
|---|---|---|---|---|---|
| $Fix_{basic}$ | 0.99 | 36.69 ±22.58 | 49.77 ±43.56 | -235.6 | 0.94 |

The analysis of the fixations number.

Although it is clear that the fixation number is an excellent parameter for classifying participants in our study group, it should not be used on its own because it can be affected by other pathological situations. For example, a participant with poor visual acuity may need to do many fixations as well. Tiredness may also affect the number of fixations, although frequently it is possible to spot tired participants from the optical scan-path that they follow as they read.

**Not fixated words $F^0$, Singly Fixated Words $F^1$, Multiply Fixated Words $F^{\geq 2}$:** The first word specific set of parameters to be evaluated consists of the number of not fixated words $F^0$, the number of singly fixated words $F^1$ and the number of multiply fixated words (more than one fixation) $F^{\geq 2}$. However, since $F^1 = total\ words - F^0 - F^{\geq 2}$, only $F^0$ and $F^{\geq 2}$ are used in the evaluation of the TRS. The results are summarized in Table 6.

**Table 6. Not fixated, singly fixated and multiply fixated words for the basic text.**

| | $A_{ROC}$ | $\left|\overline{F_{Test}^{Contr} - F_{Retest}^{Contr}}\right|$ ±SD | $\left|\overline{F_{Test}^{Dysl} - F_{Retest}^{Dysl}}\right|$ ±SD | $\overline{F_{Contr}} - \overline{F_{Dysl}}$ | $r_F$ |
|---|---|---|---|---|---|
| $F^0$ | 0.91 | 8.77 ±7.70 | 6.27 ±6.22 | 18.0 | 0.74 |
| $F^1$ | 0.90 | 8.31 ±9.16 | 6.64 ±6.17 | 17.0 | 0.71 |
| $F^{\geq 2}$ | 0.98 | 11.54 ±7.96 | 7.27 ±5.83 | -35.0 | 0.89 |

The analysis of the number of not fixated words $F^0$, the number of singly fixated words $F^1$ and the number of multiple fixated words (more than one fixation) $F^{\geq 2}$.

**Number $X_L$ of words with L letters causing gaze difficulty** (Cumulative probability above 90% in the control population word gaze duration distribution): The results are summarized in Table 7.

**Table 7. Gaze difficulty based on word length for the basic text.**

| | $A_{ROC}$ | $\left|\overline{X_{Test}^{Contr} - X_{Retest}^{Contr}}\right|$ ±SD | $\overline{\left|X_{Test}^{Dysl} - X_{Retest}^{Dysl}\right|}$ ±SD | $\overline{X_{Contr}} - \overline{X_{Dysl}}$ | $r_X$ |
|---|---|---|---|---|---|
| $X_{\leq 3L}$ | 0.93 | 3.62 ±3.20 | 6.91 ±3.95 | -20.0 | 0.89 |
| $X_{4,5L}$ | 0.93 | 1.92 ±1.75 | 3.73 ±3.30 | -8.8 | 0.80 |
| $X_{6,7L}$ | 0.96 | 1.38 ±1.26 | 2.91 ±1.63 | -11.1 | 0.93 |
| $X_{\geq 8L}$ | 0.97 | 2.38 ±3.33 | 6.32 ±4.64 | -19.0 | 0.89 |

The analysis of gaze difficulty for all words in relation with the number of letters of the words in the text.

**Total Reading Score**

The total reading score of a participant is the probability that they belong to the control group, given the values of the parameters retained that they yield for the particular text, i.e. $p(control/\underline{N})$, where $\underline{N}$ is the vector of all 12 parameters used. The way we compute this probability $p$ is given in the Appendix. If the test score is above 0.5 the individual is classified as control, while if the test score is below 0.5 the individual is classified as reader with dyslexia in the particular text.

The TRS was circularly validated using the following procedure: We trained our TRS classifier using data from 68 participants (control and dyslexic) leaving 1 individual out each time, and we asked from the program to classify this individual as dyslexic or not. This way we achieved success of 94.2% correct classification (sensitivity 93.8%, specificity 94.6%). We see that the test performed well under circular validation despite the considerable age span of the population examined (4 years). The test score of all the individuals in the text is shown in the left panel of Figure 4. Note that, without circular validation, perfect classification was achieved.

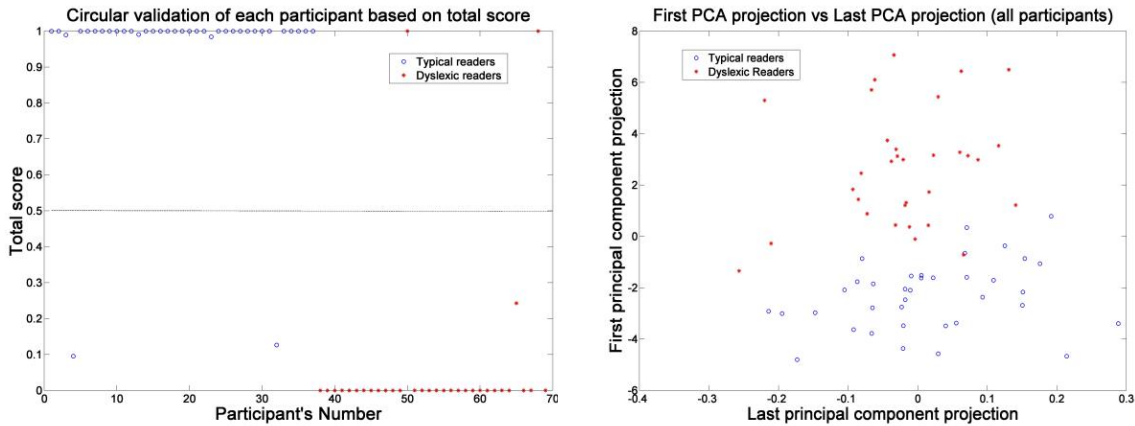

**Figure 4:** *Left:* Circular validation based on the TRS of the participants. A score above 0.5 (dash line) classifies the participant as typical reader, while a score below 0.5 classifies the participant as potentially dyslexic. *Right:* Principal component analysis is carried out on the normalized ((P-μ)/σ) vector of all 12 useful parameters of RADAR. The first principal component (maximum variance) vector projection is plotted against the last principal component (minimum variance) vector projection. The separability of readers with dyslexia from typical readers is clear.

## TEXT DEPENDENCE OF THE RESULTS

Apart from the basic text that was given to all individuals during the assessment, there was also another text, an easier text that was given after the basic text, for experimental purposes. The meaning of the text was quite simple in order to match the lower ages of the participants. Twelve children (5 control, 7 dyslexic) were rejected in this experiment, using the same exclusion criteria as in the basic text, i.e. a participant could be excluded due to unreliable eye-movement recording or lack of cooperation with the experimenters. Overall, 66 children (34 girls, 32 boys, 35 control, 31 dyslexic) read this text, all native Greek speakers. The text was written by a special education teacher in order to be appropriate for the participants' age group (8.5-12.5 years) and had 143 words, most of them one and two-syllable. Text statistics for the easy test are shown in Table 8.

**Table 8. Text statistics for the easy text.**

| | |
|---|---|
| Total word count: | 143 |
| Unique words: | 95 |
| Total number of characters: | 752 |
| Number of characters without spaces: | 610 |
| Average characters per word: | 4.26 |
| Average syllables per word: | 1.70 |
| Sentence count: | 10 |
| Max sentence length (words): | 3 |
| Min sentence length (words): | 10 |

Various text metrics for the easy text.

The same parameters as in the basic text were analyzed, although two of them, the number of refixations R and the number of not fixated words $F_0$, marginally fail the requirement that the control-dyslexic difference is at least twice the test-retest difference of both populations.

**Fixation Duration:** The parameter τ gave the data shown in Table 9 for the easy text.

**Table 9. Parameter τ for the easy text.**

| | $A_{ROC}$ | $\overline{\left|\tau_{Test}^{Contr} - \tau_{Retest}^{Contr}\right|}$±SD | $\overline{\left|\tau_{Test}^{Dysl} - \tau_{Retest}^{Dysl}\right|}$±SD | $\overline{\tau_{Contr}} - \overline{\tau_{Dysl}}$ | $r_\tau$ |
|---|---|---|---|---|---|
| $\tau$ easy | 0.87 | 16.11ms ±16.22 | 39.32ms ±27.23 | -93.1ms | 0.86 |

The analysis of the parameter τ for the easy text.

**Saccade Length:** The data obtained is summarized in Table 10.

**Table 10. Saccade length for the easy text.**

|  | $A_{ROC}$ | $\overline{\left\| q_{Test}^{Contr} - q_{Retest}^{Contr} \right\|}$ ±SD | $\overline{\left\| q_{Test}^{Dysl} - q_{Retest}^{Dysl} \right\|}$ ±SD | $\overline{q_{Contr}} - \overline{q_{Dysl}}$ | $r_q$ |
|---|---|---|---|---|---|
| $q_{25\%}$ | 0.96 | 8.37pxl ±7.24 | 5.09pxl ±5.40 | 28.8pxl | 0.90 |
| $q_{50\%}$ | 0.97 | 10.46pxl ±8.52 | 7.20pxl ±5.96 | 41.5pxl | 0.92 |
| $q_{75\%}$ | 0.94 | 14.80pxl ±9.73 | 9.40pxl ±6.82 | 53.5pxl | 0.95 |

The analysis of the saccade length for the easy text.

**Refixations:** The results for the number of refixations $R$ is given in Table11.

**Table 11. Refixations for the easy text.**

|  | $A_{ROC}$ | $\overline{\left\| R_{Test}^{Contr} - R_{Retest}^{Contr} \right\|}$ ±SD | $\overline{\left\| R_{Test}^{Dysl} - R_{Retest}^{Dysl} \right\|}$ ±SD | $\overline{R_{Contr}} - \overline{R_{Dysl}}$ | $r_R$ |
|---|---|---|---|---|---|
| $R_{easy}$ | 0.77 | 13.64 ±11.08 | 11.46 ±12.47 | -25.1 | 0.84 |

The analysis of the refixations for the easy text.

**Fixation Number**: The results for the number of fixations $Fix$ is summarized in Table12.

**Table 12. Fixations number for easy text.**

|  | $A_{ROC}$ | $\overline{\left\| Fix_{Test}^{Contr} - Fix_{Retest}^{Contr} \right\|}$ ±SD | $\overline{\left\| Fix_{Test}^{Dysl} - Fix_{Retest}^{Dysl} \right\|}$ ±SD | $\overline{Fix_{Contr}} - \overline{Fix_{Dysl}}$ | $r_F$ |
|---|---|---|---|---|---|
| $Fix_{easy}$ | 0.89 | 45.36 ±45.44 | 33.64 ±27.44 | -123.1 | 0.91 |

The analysis of the fixations number for the easy text.

**Not fixated words $F^0$, Singly Fixation Words $F^1$, Multiple Fixated Words $F^{\geq 2}$:** The results are summarized in Table 13.

**Table 13. Not fixated, singly fixated and multiply fixated words for the easy text.**

|  | $A_{ROC}$ | $\overline{\left\| F_{Test}^{Contr} - F_{Retest}^{Contr} \right\|}$ ±SD | $\overline{\left\| F_{Test}^{Dysl} - F_{Retest}^{Dysl} \right\|}$ ±SD | $\overline{F_{Contr}} - \overline{F_{Dysl}}$ | $r_F$ |
|---|---|---|---|---|---|
| $F^0$ | 0.89 | 5.92 ±5.63 | 4.95 ±4.13 | 9.0 | 0.73 |
| $F^1$ | 0.78 | 10.15 ±11.62 | 7.45 ±5.39 | 6.5 | 0.80 |
| $F^{\geq 2}$ | 0.89 | 15.77 ±12.28 | 6.32 ±5.08 | -37.5 | 0.89 |

The analysis of the number of not fixated words $F^0$, the number of singly fixated words $F^1$ and the number of multiple fixated words (more than one fixation) $F^{\geq 2}$ for the easy text.

**Number $X_L$ of words with L letters causing gaze difficulty:** The results are summarized in Table 14.

**Table 14. Gaze difficulty based on word length for the easy text.**

|  | $A_{ROC}$ | $\overline{\left| X_{Test}^{Contr} - X_{Retest}^{Contr} \right|} \pm$SD | $\overline{\left| X_{Test}^{Dysl} - X_{Retest}^{Dysl} \right|} \pm$SD | $\overline{X_{Contr}} - \overline{X_{Dysl}}$ | $r_X$ |
|---|---|---|---|---|---|
| $X_{L \leq 3}$ | 0.93 | 2.91 ±2.77 | 4.55 ±4.00 | -15.5 | 0.88 |
| $X_{L=4}$ | 0.89 | 1.64 ±2.11 | 3.77 ±1.88 | -7.8 | 0.85 |
| $X_{L=5}$ | 0.96 | 1.36 ±1.63 | 2.23 ±1.74 | -8.0 | 0.90 |
| $X_{L \geq 6}$ | 0.93 | 1.91 ±1.22 | 2.55 ±1.74 | -9.5 | 0.91 |

The analysis of gaze difficulty of words, in relation with the number of letters, for the easy text.

**Total Reading Score**

The circular validation was performed using 65 participants to train our test score classifier and one individual each time for testing. This way we achieved success of 87.9% correct classification, which was slightly lower than the success of the other text. This can be attributed to the fact that the second text was easier to read even for part of the dyslexic population, blurring the borderline between the two populations.

## UNIVERSALITY OF RADAR: EXPLORATORY RESULTS IN THE ENGLISH LANGUAGE

One might wonder whether the results were specific for the Greek language. Thus, in parallel with the study described here, there was another study by our research team in Wales, applying the RADAR method in English speaking population. The specifications of the experiment were similar, although the population was smaller: native English speakers were divided in two groups, typical readers and readers with dyslexia. They were all assessed for silent reading with two English texts.

Thirty-one children participated in the English study (16 girls and 15 boys) between 9.0 and 11.5 years of age. Eleven of them (7 girls, 4 boys) were diagnosed as readers with dyslexia by the "Tomorrow's Generation School" [39] which is a private learning center for dyslexic children and young people in Cardiff, Wales. These children constitute the dyslexic group of the English study. The remaining 20 children (9 girls, 11 boys) were not diagnosed to have reading difficulties by the Welsh school system. These children constitute the control group of the English study. All children were native English speakers and were assessed with the RADAR method in Tomorrow's generation School and in St. Athans Primary School in Cardiff, Wales.

In the English study, participants were asked to silently read the two texts. Both texts were written by a special education needs teacher in order to be appropriate for the participants' age. The first text, the "easy" text, was 141 words long, with simple text meaning, syntax and with short word length, in order to match the lower ages of the participants. The second text, the "advanced" text, had 180 words, many of them multi-syllable, and the meaning of the text was more complicated. See Table 15 for the statistics of both texts that were used.

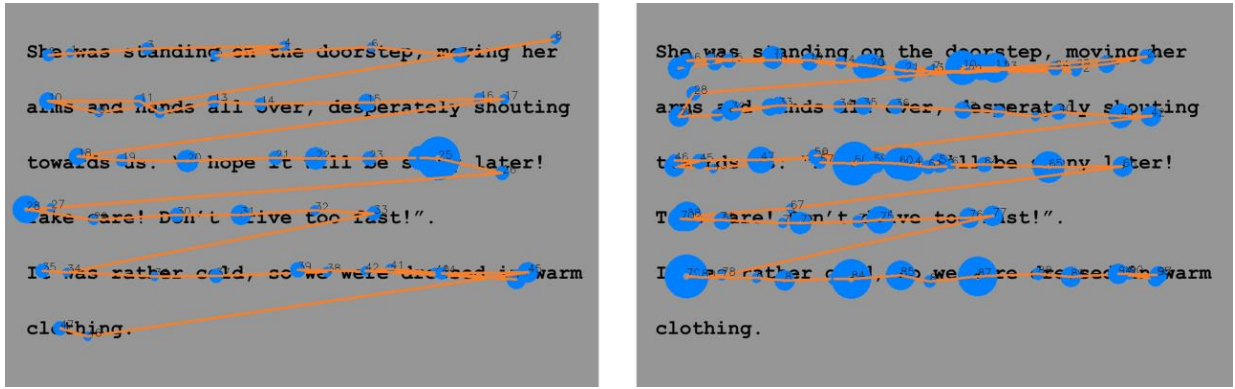

**Figure 5:** A reading path from a typical (normal) reader (_left_) and from a reader with dyslexia (_right_). Clearly the dyslexic reader exhibits longer fixations duration, shorter saccadic movements, many regressive movements and longer reading time than the typical reader.

**Table 15. Text statistics for the English texts.**

| Parameter/Text | Easy test | Advanced text |
|---|---|---|
| Total word count: | 141 | 180 |
| Unique words: | 76 | 126 |
| Total number of characters: | 759 | 1047 |
| Number of characters without spaces: | 622 | 867 |
| Average characters per word: | 4.41 | 4.81 |
| Average syllables per word: | 1.43 | 1.63 |
| Sentence count: | 18 | 28 |
| Max sentence length (words): | 9 | 10 |
| Min sentence length (words): | 3 | 1 |

Various text metrics for both easy and advanced text use in the English study.

The analysis performed is identical to the non word-specific analysis of the Greek study. Word-specific analysis was not possible due to the small size of the population, which did not allow us to analyze more parameters. The fixation duration parameter $\tau$ gave $A_{ROC}=0.85$ for the advanced test and $A_{ROC}=0.74$ for the easy text. For the English speaking population, the parameter that gave the highest discriminability was the number of refixations, which gave $A_{ROC}$=0.92 for both the advanced and the easy text. This we attribute to the fact that readers of non-transparent languages tend to exhibit higher number of refixations.[18] The mean saccade length parameter gave $A_{ROC}$=0.85 for the advanced text and $A_{ROC}$=0.76 for the easy text.

Overall, English dyslexic readers had longer fixations, shorter saccade lengths, more refixations and more fixations compared with non-dyslexic readers. These findings are in agreement with other studies for English dyslexic readers.[2,40] The fact that our measurements are in agreement with prior studies increases confidence in our approach. In agreement with other studies[18] is also the fact that higher numbers of refixations are associated with non-transparent languages than with transparent. Three of the dyslexic individuals examined stood out in that their non word-specific parameters matched the control populations' values except for the number of refixations, which was strongly in the range of readers with dyslexia. Readers with dyslexia with similar characteristics did not appear in the Greek study. This may have to do with the non-transparency of the English language, or it may have to do with the way readers with dyslexia are diagnosed in the two languages. We hope to clarify this in the future. These observations point out the possibility that quantitative eye movement measurements will open a window for doing more refined reading analysis, potentially helping to define or better characterize sub-categories of dyslexia.

The results above strengthen the team's belief that the RADAR method can be used as a universal screening tool, given that it can calibrated in each language separately. The next major step would be a large-scale English study to further evaluate the reading characteristics of native English speakers.

## LIMITATIONS

It is not by chance that most eye-tracking studies during reading have been performed on adults.[2,17,41] By comparison, studies on children remain few.[1,11] One reason for this is certainly that most studies target the cognitive tasks underlying reading rather than developmental dyslexia. However, another issue is the difficulty in recruiting children capable of cooperating properly in order to accumulate accurate and reliable data.[11] During the data acquisition the participant has to stay as still as possible and follow a specific set of instructions, making this process challenging even for some adults. An important component of our method was the development of a child-friendly custom eye-tracker and special eye-tracking software that allowed us to obtain accurate and precise measurements in most children examined.

A complicating factor for our analysis is that eye-movement parameters change during development: reading speed increases, fixation durations decrease, longer saccadic movements occur, etc.[18] Hazel and Holly[11] collected eye-movement characteristics observed across ages during reading in a variety of studies of normally developing children. In the present study, even though eye movement characteristics change during development, we analyzed all the children (ages 8.5-12.5) as one group. The reason is that we did not have enough participants for more refined age analysis. It should be noted that despite the 4-year age span it was still possible to discriminate participants with dyslexia from controls reliably and efficiently. Fig. 6 shows how extracted, non word-specific, parameters varied with age in our population. As our database increases, more fine-grained analysis will be possible, further improving the performance of the RADAR measure.

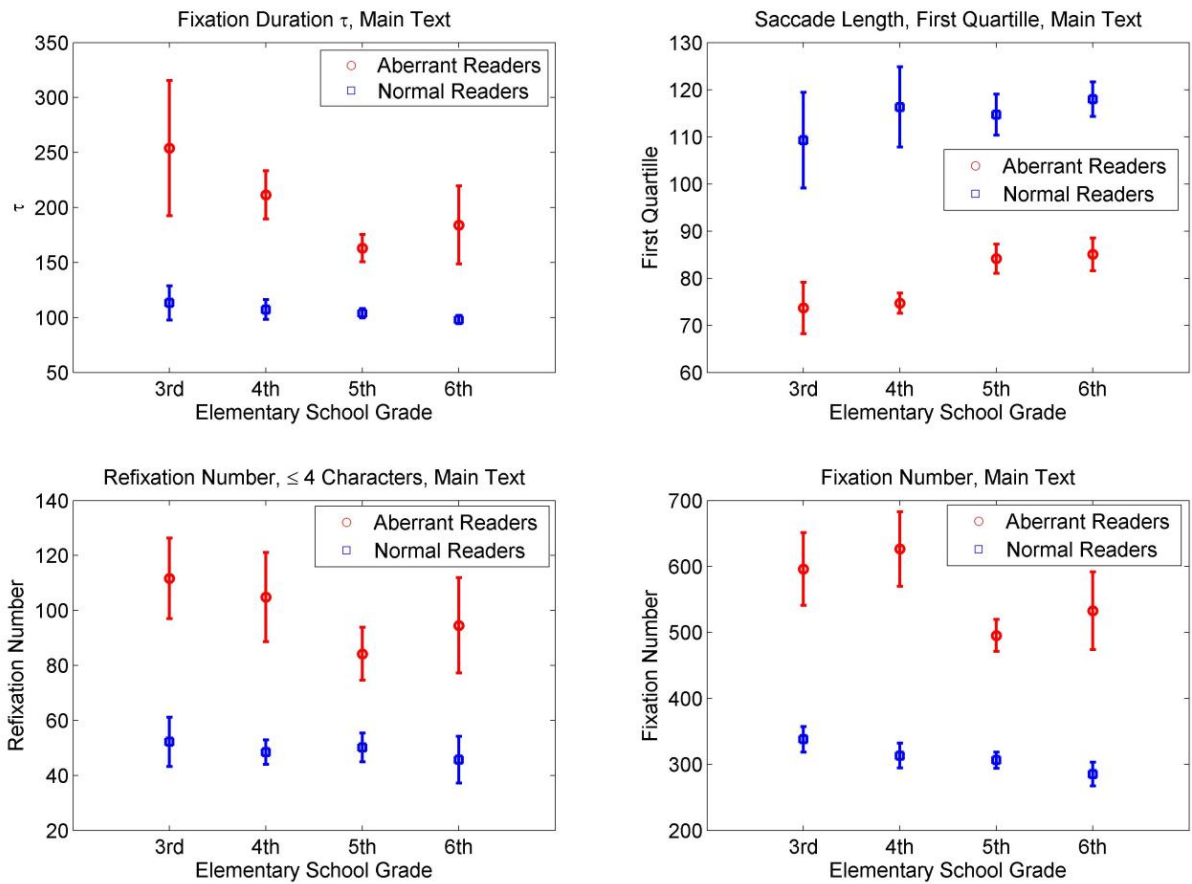

**Figure 6:** Non word-specific parameter dependence on age. Error bars represent standard error of the mean. Although the population size for each grade is small, the trends of the parameters are evident: mean fixation duration (τ) drops with age, saccade length (first quartile plotted here in pixels) increases with age, number of refixations decreases with age, and number of fixations decreases with age. Note that the two populations (dyslexic and typical readers) are clearly separable throughout the entire four-year age span.

A current technical limitation is that eye-tracking requires from the individuals that are assessed to have their head as still as possible during calibration. We use a chin rest to help with this but keeping the head still remains somewhat difficult for young children. Nevertheless, in our group of 8.5-12.5 age participants, it was not difficult to obtain high quality eye tracking. Only 9/78 participants were excluded (blind to the diagnosis) on the basis of either poor eye tracking or lack of cooperation. We are currently trying to eliminate the need for a chin rest through the implementation of sufficiently accurate head tracking.

It should be stressed that RADAR provides information to the therapist that is largely complementary to the information provided by currently used tests. Quantitative information about how individual subjects scan specific words or sentences is likely to be valuable for designing individualized therapeutic strategies and monitoring recovery. RADAR is not designed to evaluate reading accuracy, which requires loud reading, nor to evaluate comprehension, since this can be done easily with currently used tests. Nevertheless, the parameters evaluated by RADAR are sufficient to screen children for dyslexia efficiently.

## DISCUSSION

The present study demonstrates that the RADAR method can separate effectively dyslexic from non-dyslexic readers, based on a series of eye-tracking parameters obtained during the silent reading of a standard text. Although it is clear that dyslexia is not caused by oculomotor deficits[42], readers with dyslexia do have different patterns of eye movements during reading compared to non-dyslexic readers.[18,43,44] We measured and compared eye movement patterns of dyslexic and non-dyslexic readers 8.5-12.5-year-old. Our results corroborate the results of prior studies that eye movement patterns are different between dyslexic and non-dyslexic readers[18,43,44], but go further in several respects. Specifically, i) we evaluate the power of different eye tracking parameters in discriminating among typical and dyslexic readers, ii) we assess the stability of these parameters under retesting, and iii) we combine parameters that have high discriminability and stability into a score that can classify 8.5-12.5-year-old Greek readers as typical or atypical with high sensitivity (93.8%) and specificity (94.6%).

Specific results obtained depend in principle on the reading text chosen. The reading task that RADAR uses consists of a short text, selected by a special educational needs teacher to be appropriate for the diagnosis of dyslexia (see Table 1 for text statistics). The text size was designed to minimize participant's fatigue or lack of cooperation, while yielding a rich and objective set of eye tracking data that can be used for classifying participants as being at-risk for dyslexia. Eye-tracking evaluation was performed during the silent reading of this text. The mode of reading is important, since eye movements differ when reading silently versus aloud [45], perhaps because in the latter case participants have both to decode and articulate the text. Silent reading was chosen as it is the principal mode of reading for learning after grade 3. It is important to note that results obtained were replicated using two different types of reading text, one easy and one difficult (see Table 8 for easy text and Table 1 for difficult (basic) text), strengthening the validity of our conclusions. Further optimization of the text used for evaluating eye-tracking parameters can potentially lead to improved power for discriminating dyslexic from normal readers in the future.

It is of interest to briefly consider the discriminating power provided by the extracted RADAR eye-tracking parameters. RADAR parameters, such as fixation duration, saccade length, number of refixations etc., provide information that is not accessible to speech pathologists carrying out a standard printed test. In the Results section, we have shown that these parameters are reproducible under retest and have discriminating potential as seen from the area under their ROC curve. The optimal choice of threshold of discriminability for each parameter depends on the particular demands of sensitivity and specificity. A reasonable choice is the threshold that corresponds to the point in the ROC curve that is closest to the point (0,1) (see Fig. 7).

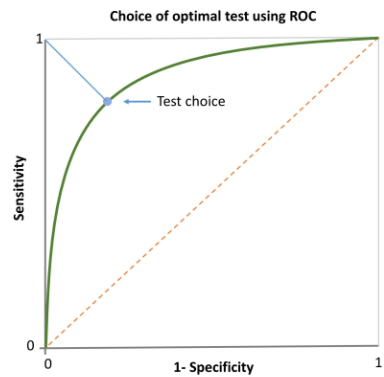

**Figure 7:** How to choose the best possible test using ROC

Table 16 lists the sensitivities and specificities of the extracted RADAR parameters for this choice of threshold. Several parameters listed in Table 16 can rapidly and accurately signal deviations from the typical reading population.

**Table 16. Sensitivity, specificity and threshold values for both texts.**

| Parameter | Basic Text | | | Easy Text | | |
|---|---|---|---|---|---|---|
| | Sensitivity | Specificity | Thresh | Sensitivity | Specificity | Thresh |
| Exponential decay $\tau$ (ms) | 0.88 | 0.86 | 123.2 | 0.77 | 0.91 | 151.4 |
| Saccade Length $q_{25\%}$ (px) | 0.91 | 0.95 | 95.0 | 0.94 | 0.91 | 95.3 |
| Saccade Length $q_{50\%}$ (px) | 0.91 | 0.92 | 123.7 | 0.94 | 0.91 | 124.3 |
| Saccade Length $q_{75\%}$ (px) | 0.91 | 0.86 | 168.7 | 0.87 | 0.91 | 154.5 |
| Refixations (R) | 0.75 | 0.84 | 67.1 | 0.65 | 0.80 | 52.1 |
| Fixations Number (Fix) | 0.97 | 0.97 | 384.2 | 0.87 | 0.77 | 277.1 |
| Not fixated words $F^0$ | 0.84 | 0.78 | 17.1 | 0.84 | 0.83 | 10.1 |
| Singly fixated words $F^1$ | 0.84 | 0.84 | 60.1 | 0.68 | 0.80 | 50.1 |
| Multiply fixated Words $F^{\geq 2}$ | 0.94 | 0.92 | 100.1 | 0.84 | 0.80 | 75.2 |
| Words (X) with (L) letters causing difficulty $X_{L \leq 3}$ | 0.88 | 0.86 | 12.0 | 0.94 | 0.80 | 8.0 |
| $X_{L=4,5L}$ ($X_{L=4}$ for easy text) | 0.81 | 0.89 | 4.0 | 0.77 | 0.86 | 5.0 |
| $X_{L=6,7}$ ($X_{L=5}$ for easy text) | 0.97 | 0.89 | 5.0 | 0.94 | 0.94 | 4.0 |
| $X_{L \geq 8}$ ($X_{L \geq 6}$ for easy text) | 0.91 | 0.92 | 10.0 | 0.94 | 0.80 | 4.0 |

The sensitivity and specificity values with the corresponding threshold, for all the parameters of the analysis.

The RADAR screening decision is made using a total reading score (TRS) based on these parameters. TRS represents the probability that the reader is non-dyslexic given the measured eye-tracking parameter vector, consisting of the parameters: $\tau$, $q_{25\%}$, $q_{50\%}$, $q_{75\%}$, $R$, Fix, $F^0$, $F^{\geq 2}$, $X_{L \leq 3}$, $X_{L=4,5}$, $X_{L=6,7}$, $X_{L \geq 8}$ shown in Table 16, which is taken to obey a multi-variate Gaussian distribution (see Appendix). Parameter $F^1$ is omitted because it is fully dependent on $F^0$ and $F^{\geq 2}$. Under circular validation, the sensitivity and the specificity of the TRS are 93.8% and 94.6%. Note that without circular validation, the performance of TRS is near perfect: we evaluated TRS over the entire population of participants in order to compare its sensitivity and specificity with that of the individual parameters listed in Table 16. As expected, TRS performed better than individual parameters, giving sensitivity and specificity 1 for the basic text, and sensitivity 1 and specificity 0.94 for the easy text. Note that these numbers were obtained without circular validation, which is why they are better than the numbers reported in the abstract. The fact that the TRS classifier is near perfect both with and without circular validation, suggests that as the training set becomes larger its performance on subjects it was not trained on will continue to improve. After all, if the sample is large, the exclusion of one testing participant (as done for circular validation) is not expected to influence significantly the classifier construction.

It is important to note that our main observations were supported in both Greek and English speaking populations. This is an important validating test demonstrating the generalizability of the RADAR method from the transparent Greek language to the non-transparent English language. Greek dyslexic readers exhibited longer fixation durations, shorter saccadic movements, higher number of backward refixations and higher number of overall fixations than typical readers. Similar effects have been observed in other transparent languages, which exhibit relatively simple grapheme-phoneme relations, for example Italian and German.[18,43,44] A high number of fixations for longer words and a high frequency of short saccadic eye movements were two of the most prominent reading characteristics for Greek dyslexic readers. From our exploratory study for the non-transparent English language, we found that eye-tracking characteristics of readers with dyslexia were similar, though they had different discriminability. Interestingly, in the English language the parameter with the highest discriminability was the number of short backward refixations as compared to the Greek language for which it was the overall number of fixations. It will be interesting to follow these preliminary observations in larger studies in the future.

Apart from accurately classifying participants as typical or atypical (possibly dyslexic) readers, the RADAR method provides valuable data that can be used to study the reading strategy that each individual employs. Such information can be important to the specialist for identifying words or phrases that cause difficulty in reading to the individual. For example, the word-specific parameters, provide information about the way a participant reads specific words in silent reading. Refixations and long gaze duration on specific words indicate a difficulty in decoding the word and appear frequently in many readers with dyslexia. This analysis can help reading specialists select particular words that cause difficulty to the individual participant. We plan in a future study to associate different eye tracking parameters with specific reading difficulties. We predict that, as we accumulate more participants and discriminating power increases, it may well become possible to identify subclasses of dyslexia that exhibit distinct reading abnormalities among individuals that have a positive RADAR result.

A screening method based on objective data collected by the clearly defined and simple protocol that RADAR uses, will potentially be helpful for clarifying issues that may arise as a result of the specific definition of dyslexia used. Although most experts subscribe to the definition of dyslexia given in Lyon et al.[3], there is still no universally accepted definition of dyslexia. As stated in the book[46] "Dyslexia in the Primary Classroom" by Hall W., the lack of a universally accepted definition of dyslexia is partly due to a lack of clarity in understanding the relationship between reading and language and a failure to identify stable correlates of dyslexia. An added complication is that the definition and diagnostic criteria of dyslexia naturally differ among countries and languages. It would therefore be helpful to have a method that collects quantitative data on the reading strategy subjects use, with parameters that can be measured in any language and can be compared across languages, though standardized in each language separately. RADAR is such a method.

The value of RADAR as a screening method is clarified by comparing the prior probability of a Greek school child 8.5-12.5 years of age to be dyslexic with the posterior probability to be dyslexic after a positive RADAR result. The international prevalence of dyslexia is believed to be at least 10%, although numbers vary according to the severity of dyslexia required to characterize a subject as dyslexic [47]. Hence it is reasonable to take the prior probability of a child to be dyslexic to be $P_{prior}^D = 0.1$. Hence the prior ratio of probabilities is $r_{prior}^+ = \frac{P_{prior}^D}{P_{prior}^N} = \frac{1}{9} = 0.111$. Since the (circularly validated) sensitivity of the RADAR total score is 93.8% and its specificity is 94.6%, the likelihood ratio $L^+ = (\text{sensitivity}/(1 - \text{specificity}) = 93.8/(1 - 94.6) = 17.37$. The posterior ratio of probabilities after a positive RADAR result is $r_{post}^+ = \frac{P_{post}^D}{P_{post}^N} = L^+ * r_{prior}^+ = 1.93$. This gives us that $P_{post}^D = \frac{1.93}{1.93+1} = 0.66$, hence the probability that a child positively tested under RADAR is dyslexic is 66%, given that the child examined satisfies the selection criteria described in Methods. Hence the value of screening with RADAR is that a positively tested child, instead of having the general population 10% probability of being dyslexic, has the much higher 66% probability of being dyslexic. In the case of a negative RADAR result, $r_{prior}^- = \frac{P_{prior}^N}{P_{prior}^D} = 9$, $L^- = \frac{\text{specificity}}{1-\text{sensitivity}} = \frac{94.6}{1-93.8} = 15.26$, hence $\frac{P_{post}^N}{P_{post}^D} = L^- * r_{prior}^- = 137.34$. This means that $P_{post}^N = \frac{137.34}{137.34+1} = 0.993$. Hence a negatively tested child under RADAR, instead of the general population probability 90% of being non dyslexic has the posterior probability 99.3% of being non dyslexic.

One further advantage of the RADAR method is that it is difficult to be manipulated by people who pretend to be dyslexic, to gain favorable treatment (in examinations for example). Text that has been skipped by the individual can be easily identified, as can text that has been read more than one time. Just by analyzing the reading path, one can observe potentially deliberate delays in reading introduced by the individual. It is important to note that the results we obtained were excellent despite the fact that we grouped participants with ages from 8.5-12.5 years together. In future studies, larger number of participants will allow us to narrow the age interval, further increasing the power of our analysis by comparing participants with tighter age-appropriate controls.

Another advantage of the RADAR method is that it can be applied to monitor the effectiveness of a treatment to individuals with a specific reading disorder. The RADAR method can be re-administered at regular time intervals, based on the opinion of the specialist that administered the treatment. The outcome can be compared to the control reading population and to previous assessments of the individual. Treatment effectiveness can be gauged by how closely the participant's reading parameters approximate those of the control population. If progress is poor, then the specialist can alter the treatment.

Naturally, the RADAR method is not meant to substitute expert opinion on the diagnosis of specific reading disorders. There are many additional parameters, pertinent to dyslexia, that are not assessed by the RADAR method, such as working memory, accuracy of word decoding, and others. However, we have demonstrated here that RADAR represents a useful tool, which provides the specialist with quantifiable and objective data on reading performance. Such data would not be available to specialists without the RADAR eye tracking technology and analysis.

Finally, we believe that there is ample space for further development of the RADAR method. We have not yet used eye-tracking parameters to evaluate traditional printed dyslexia tasks, for example reading of pseudo-words or similar sounding words. This was avoided in order to keep the test short and easy. Ideally, an optimized test that would include both appropriately chosen text and specific reading tasks would be able not only to detect dyslexia but to specify sources of difficulty specific to the individual examined. Whatever the future may be, RADAR yields abundant quantitative and objective data that can be used to evaluate reading in a fast, reliable, non-invasive way, helping to identify children with probable dyslexia. Furthermore, RADAR is easy to implement and low cost, making it appropriate for large-scale screening of school-age children populations.

Acknowledgments: Special thanks to Joanna Christodoulou, PhD, Human Development and Psychology, Nicholaos Makaronas, MD, Child Psychiatrist, Panagiotis Dimitrakopoulos, MD, Ophthalmologist, Thomas Karantinos, PhD, Neuroscientist, Maria Rousochatzaki, MSc., Speech Pathologist

# APPENDIX

The results obtained, were collected in a vector $\underline{N}$, and combined as follows to obtain the final reading score: First all the parameters in $\underline{N}$ were normalized to $\underline{Z}$ by subtracting the mean and dividing by the standard deviation. Then, the covariance matrices $C_{contr}$, $C_{dysl}$ of the parameters $\underline{Z}$ were obtained for the control and the dyslexic populations respectively. Multivariate Gaussian distributions are generated for both the control and the dyslexic populations using these covariance matrices:

$$p(\underline{z}/control) = \frac{1}{\sqrt{2\pi \det(C_{contr})}} \exp(-\frac{1}{2}\underline{z}^T c_{contrl}^{-1}\underline{z}),$$

$$p(\underline{z}/dyslexic) = \frac{1}{\sqrt{2\pi \det(C_{dysl})}} \exp(-\frac{1}{2}z^T c_{dysl}^{-1}\underline{z}),$$

These are taken to be the distributions of the vector of normalized parameters $\underline{Z}$ in each population.

To obtain the probability that an individual is in the control group versus the dyslexic group, we use the Bayes theorem to get

$$P(control/\underline{z}) = \frac{p(\underline{z}/control)p(control)}{p(\underline{z}/control)p(control) + p(\underline{z}/dyslexic)p(dyslexic)},$$

$$P(dyslexic/\underline{z}) = \frac{p(\underline{z}/dyslexic)p(dyslexic)}{p(\underline{z}/control)p(control) + p(\underline{z}/dyslexic)p(dyslexic)},$$

where the prior probabilities of an individual being in the control or dyslexic groups are given by the fraction of typical readers versus readers with dyslexia among all participants examined.

The score of an individual is the probability $P(control/\underline{N}) = P(control/\underline{z})$ of the individual. A score above 0.5 indicates that an individual has higher probability to be in the control group, while a score below 0.5 indicates that an individual has a higher probability to be in the dyslexic group (see Fig. 4).